\documentclass{appolb}
\usepackage{epsfig}

\begin{document}
\date{\today}
\pagestyle{plain}
\newcount\eLiNe\eLiNe=\inputlineno\advance\eLiNe by -1
\title{Radiative return via electron pair production:\\
 Monte Carlo simulation of the process $e^+e^- \to \pi^+\pi^- e^+e^-$
\thanks{Presented by E. Nowak-Kubat
 at XXIX International Conference of Theoretical Physics,
 `Matter To The Deepest', Ustro{\'n}, 8-14 September 2005, Poland.
 Work supported in part by
 EC 5-th Framework Program under contract
  HPRN-CT-2002-00311 (EURIDICE network),  TARI project RII3-CT-2004-506078
 and
  Polish State Committee for Scientific Research
  (KBN) under contract 1 P03B 003 28.}
}
\author{HENRYK CZY\.Z and EL\.ZBIETA NOWAK-KUBAT
\address{Institute of Physics, University of Silesia,
PL-40007 Katowice, Poland}}
\maketitle

\begin{abstract}
 Contributions from
  the reaction $e^+e^- \to \pi^+\pi^- e^+e^-$ to the 
 pion form factor measurement via radiative return method
 are discussed basing on the results of
 a Monte Carlo generator (EKHARA).
 The generator contains contributions from 
  the initial and final state emission of a $e^+e^-$ pair
 from $e^+e^-\to\pi^+\pi^-$ production diagrams and the $\pi^+\pi^-$ 
 pair production from space-like and time-like Bhabha diagrams. 
 A detailed study 
 is performed for the $\Phi$- factory energy. 
 Tests of the generation procedure are also presented.
\end{abstract}

\section{Introduction}
 The radiative return method \cite{Zerwas}
  is a powerful tool in the 
 measurement of $\sigma (e^+e^- \to \  {\mathrm{hadrons}})$ 
 and detailed studies of hadronic interactions
 \cite{Nowak,Czyz:2004nq,Czyz:2005}.
 Very accurate knowledge of the hadronic cross section is essential
 for predictions of the hadronic contributions to $a_\mu$, the anomalous 
 magnetic moment of the muon, and to the running of the electromagnetic
 coupling from its value at low energy up to $M_Z$ (for recent reviews look
 \cite{Davier:2003pw,Jegerlehner:2003rx,Nyffeler:2004mw}).
  Due to a complicated experimental setup,
 the use of Monte Carlo (MC) event generators 
 \cite{PHOKHARAetc,actaHN},
 which include various radiative corrections \cite{PHradcor}
 is indispensable. Some more extensive analysis of that subject can be found
 also in \cite{proc}.
 The most important hadronic mode, i.e. $\pi^+\pi^-$, was recently measured 
 by KLOE \cite{KLOE1}
  by means 
 of radiative return method. In this measurement only pions
 (+ missing momenta) in the final state were observed.
  For that particular measurement there is no
 difference between photon(s) and the pair production and
 one should estimate the contribution of the process 
 $e^+e^- \to \pi^+\pi^-e^+e^-$ to the measured cross section.
 As suggested in \cite{Hoefer:2001mx} the 
contribution from the $e^+e^-$ production 
 to the $e^+e^- \to \pi^+\pi^-\gamma$ is sizable and comes mostly from the
 t-channel Bhabha--like diagrams. In this paper a Monte Carlo study
 is performed to test if this claim remains true for a realistic
  (KLOE) experimental setup.
 
\section{Monte Carlo simulation and its tests}
\begin{figure}[ht]
\begin{center}
\epsfig{file=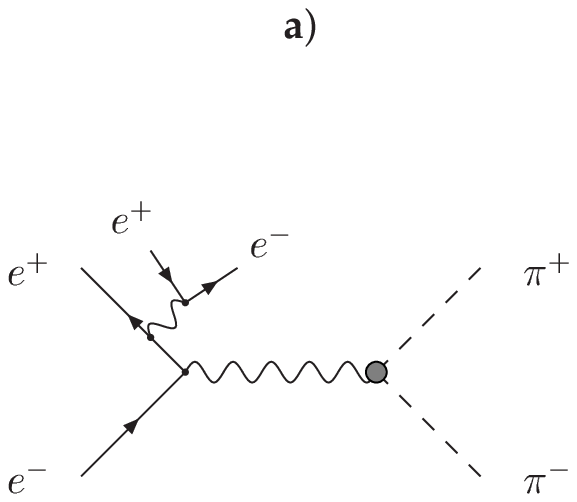,width=3.3cm,height=2.9cm}
\hskip+0.5cm
\epsfig{file=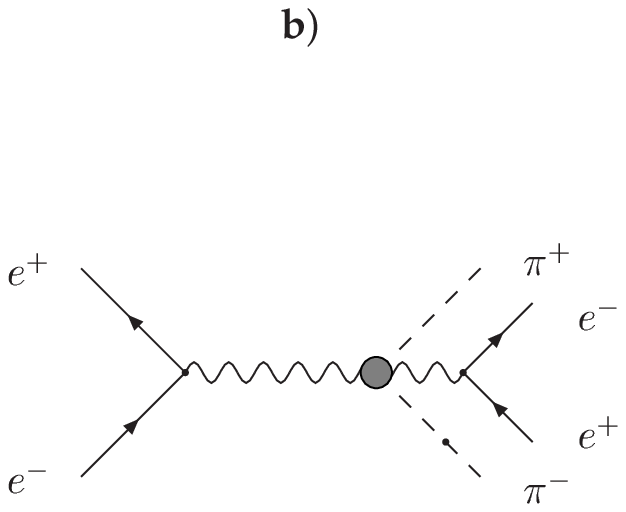,width=3.2cm,height=2.9cm}
\hskip+6cm
\epsfig{file=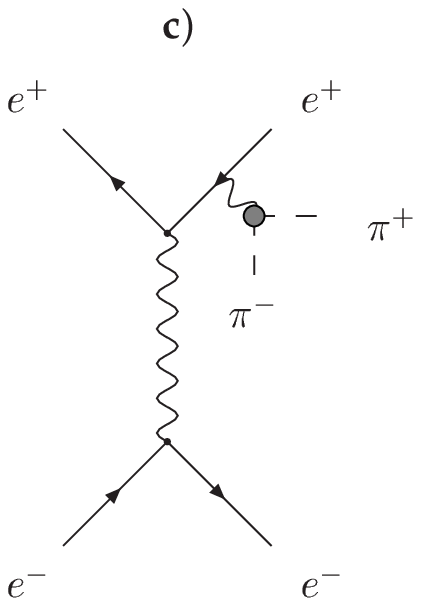,width=2.2cm,height=3cm}
\hskip+0.5cm
\epsfig{file=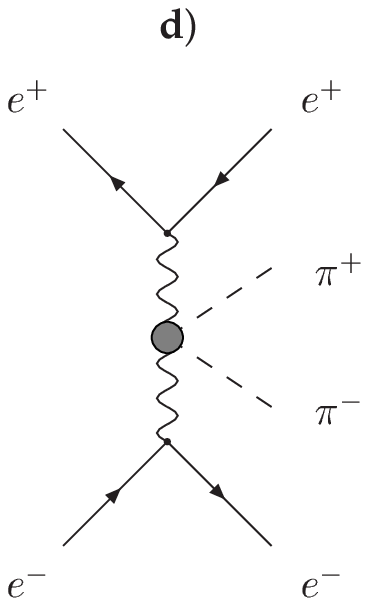,width=2cm,height=3cm}
\hskip+0.2cm
\epsfig{file=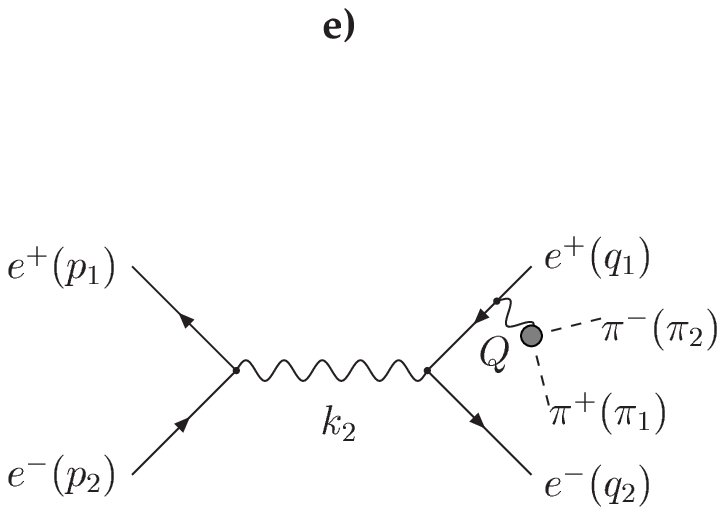,width=3.3cm,height=2.9cm}
\end{center}
\caption{Diagrams contributing to the process 
\(e^+(p_1)e^-(p_2) \to \pi^+(\pi_1)\pi^-(\pi_2)e^+(q_1)e^-(q_2)\):
 initial state electron pair emission (a),
 final state electron pair emission (b),
 pion pair emission from t--channel Bhabha process  (c),
 $\gamma^*\gamma^*$ pion pair production (d) and pion pair emission 
 from s--channel Bhabha process (e). 
}
\label{f00}
\end{figure}

 In Fig.\ref{f00} different types of diagrams
  contributing to the reaction $ e^+e^- \to \pi^+\pi^-e^+e^- $ 
 are shown schematically.
 In the present version of the Monte Carlo program
 we include all but the $\gamma^*\gamma^*$ pion pair production process
 diagrams (Fig.\ref{f00}d), which were estimated to be
 negligible for DA$\Phi$NE energy \cite{Juliet}.
 We use scalar QED  to model the FSR $e^+e^- $ pair emission
 and the $\rho$ dominance model for the $\gamma^*(\rho^*)\pi\pi$ coupling
 (see \cite{actaHN,CN_new} and \cite{Bruch:2004py} (from where the model
 of the pion form factor was implemented)  for details).
 The contribution from initial state radiation (ISR) diagrams  Fig. \ref{f00}a,
 and  final state radiation (FSR) diagrams  Fig.~\ref{f00}b was discussed 
 in details in \cite{actaHN}. 
 Because of a heavy pair emission, the contribution from diagrams presented 
 schematically in Fig.~\ref{f00}e is completely negligible for 
 any event selection used in the analysis presented below.

 We use multi--channel variance reduction method to improve efficiency of the
 generator and the generation is split into four channels, where two of them
 absorb peaks present in t--channel diagrams and other two take care
 of the s--channel peaks.
 All details will be given in a separate publication \cite{CN_new}.
 
To check correctness of the program code, we have performed a number of 
 tests of the new part of program.
 Gauge invariance of the sum of the amplitudes was checked analytically
 for set of diagrams from Fig.~\ref{f00}c,e.
 To cross check the helicity amplitudes,
 used in the program to calculate square
 of the matrix element,
  we have used also the standard trace
  method for independent calculation.
 Both results, summed over polarizations
 of initial and final leptons,
  were compared numerically scanning 
 the physical phase space. The biggest relative difference
 between the two results, which was found, was at the level of $10^{-11}$ 
 for diagrams in Fig.~\ref{f00}c
 and $10^{-23}$  for diagrams in Fig.~\ref{f00}e.
 The computer code
  has been written in quadruple precision not to lose accuracy as sever
 numerical cancellations occur even when using helicity
 amplitudes. 
\begin{figure}[ht]
\begin{center}
\hskip-0.2cm
\epsfig{file=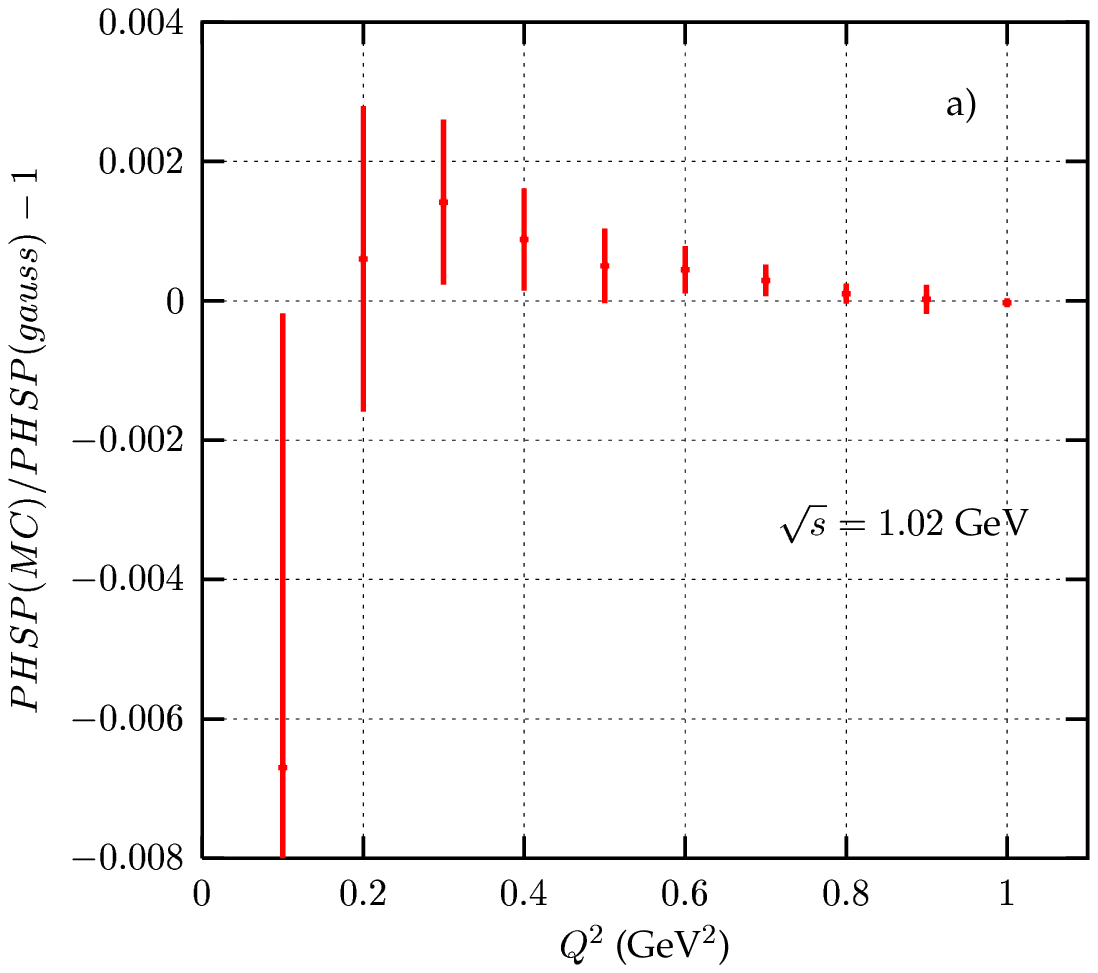,width=6.4cm,height=5.8cm}
\hskip-0.2cm
\epsfig{file=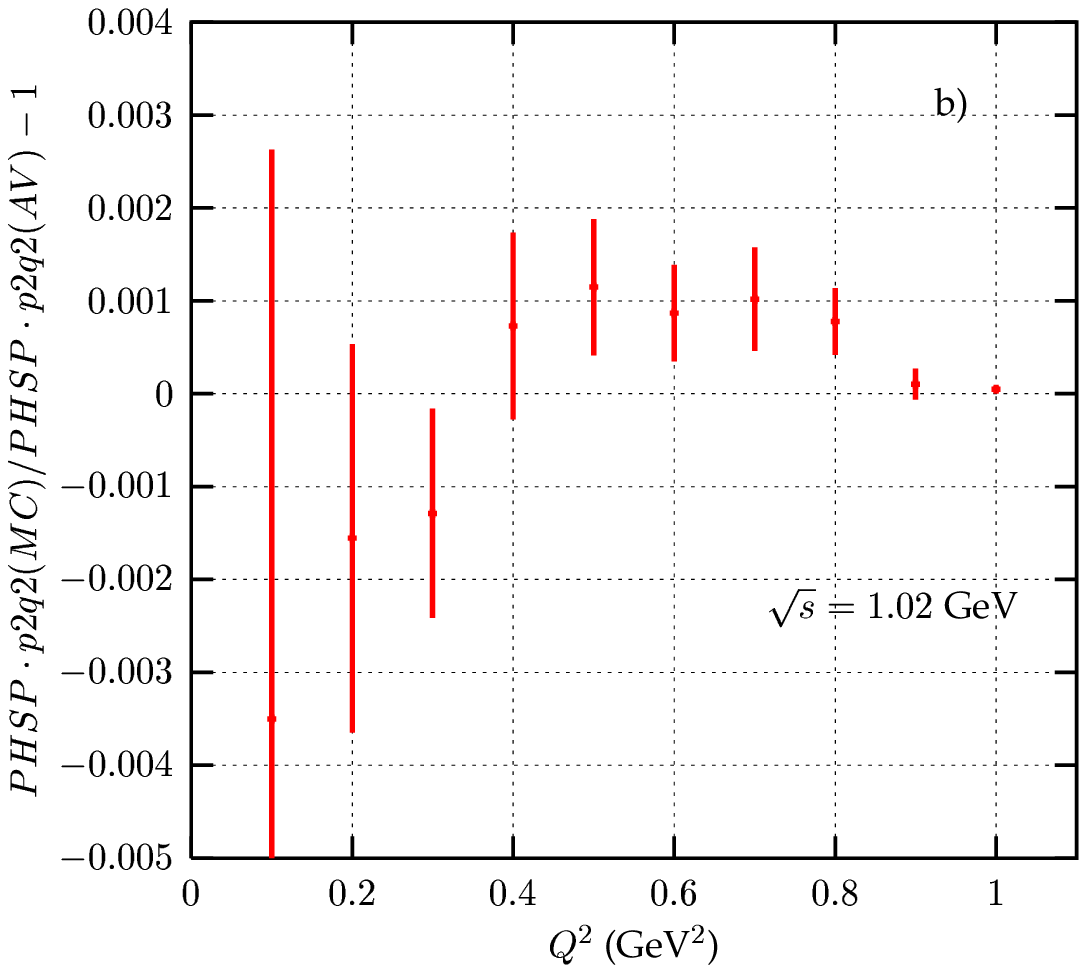,width=6.4cm,height=5.8cm}
\end{center}
\caption{Relative difference between phase space calculated by MC 
         and by gauss method. The errors come from MC integration (left).
          Relative difference between MC result
          and average for phase space $\times$ simple $|M|^2$ (right).
}
\label{phsp}
\end{figure}
The phase space volume as a function of the square of the invariant mass 
 of the pion system calculated by the Monte Carlo program was compared
 with a one dimensional Gauss integration.
  The result is presented in Fig.~\ref{phsp}a. 
  The relative difference  stays well within errors, which are 
 at most at the level of a few per mil.
 Because of the complexity of the generation we performed an additional test
 where a very simple matrix element was introduced. The matrix element
 was chosen in a way that only a four dimensional numerical integration
 was necessary.
 The matrix element
  was integrated by the Monte Carlo program with full phase space generation
 and compared with the results obtained with 4-dimensional Gauss 
 integration, where the integration region was divided into 
 ($2\cdot10^7 - 6\cdot10^7$) rectangular
 solids (or more complicated shapes at the border of integration region).
 In each rectangular solid the 8-point Gauss procedure was used recursively.
 The error of the Gauss method was estimated as the maximal difference of
 the results with different subdivisions and the central value was taken
 as a mean of the smallest and the biggest obtained values.  
  The relative difference between MC result
 and the Gauss method
 is presented in Fig.\ref{phsp}b. Again the 
 relative difference  stays well within errors, which are 
 at most at the level of a few per mil.
\begin{table}
\begin{center} 
\vspace{+0.5cm}
\begin{tabular}{|c|c|c|c|c|}
\hline
$Q^2$ [GeV$^2$] &
 R$_{t+int}$(BN) (\%) & R$_{t+int}$(MC) (\%)& R (BNK) (\%)&  R(MC) (\%)\\ 
\hline
0.09  & 8.87   & 8.06(15)    & 9.35 & 8.54(15) \\  
0.16  & 3.60   & 3.35(5)     & 4.08 & 3.84(5)  \\ 
0.25  & 1.36   & 1.46(2)     & 1.83 & 1.92(2)  \\
0.36  & 0.46   & 0.581(7)    & 0.90 & 1.028(7) \\
0.49  & 0.095  & 0.219(2)    & 0.51 & 0.639(2) \\
0.5776& 0.0056 & 0.01110(8)  & 0.40 &0.5074(8) \\
0.64  & -0.024 & 0.0658(4)   & 0.35 &0.4400(5) \\
0.81  & -0.036 & 0.01151(5)  & 0.25 &0.3001(2) \\
1.0   & -0.012 & 0.000091(1) & 0.10 &0.10916(5)\\
 \hline
\end{tabular}
\end{center}
\caption{Ratios R$_{t+int}$ and R in \% (see text for definitions) :
         BN --- results based on formulae from \cite{bervanneerv},
         BNK --- results based on formulae from \cite{bervanneerv} and
             \cite{KKKS88},
         MC --- Monte Carlo results}
\label{Tab:table1}
\end{table}

A comparison with existing analytic results \cite{bervanneerv,KKKS88}
was also performed.
In the Table \ref{Tab:table1} the ratio (denoted by R$_{t+int}$) 
of differential cross sections for t-channel 
(space-like Bhabha with radiated pion pair) plus the interference between 
t-channel and s-channel ( only the ISR pion pair radiated from  
$e^+e^-\to e^+e^-$ is included)  to the differential cross section 
of the reaction $e^+e^-\to\pi^+\pi^-\gamma$ (Born: ISR+FSR) is presented
 as a function of the invariant mass 
of the two-pion system (Q$^2$). No cuts are imposed as the analytic formulae
 exist only for that case.
 A discrepancy at the level of about 10 \% between the Monte Carlo
 and the analytic
 results can be observed in the region were the pair contribution is
 sizable and it is much larger in the region were the pair contribution
 is at the level of 1 per mil or lower. For s-channel diagrams
 we have found previously \cite{actaHN} an agreement with analytic
 results of \cite{KKKS88} at the level of 1 per mil.
 We shall observe, that even if we take
 the relative difference as the error of the MC code,
 the accuracy of the code is good enough for an estimation of the 
 contribution of the pair production to the process with the 
 photon(s) radiation.
 To estimate the complete contribution for the (unrealistic) situation
 with no cuts imposed,
 the ratio (denoted by R) of the t-
 and the s-channel pair production to the differential 
cross section of $e^+e^-\to\pi^+\pi^-\gamma$ (Born: ISR+FSR)
is also presented in Table \ref{Tab:table1} both for the MC based
 and the analytic results \cite{bervanneerv,KKKS88}.

\section{An event selection dependence of the pion pair production 
contribution to the radiative return measurement }

\begin{figure}[ht]
\begin{center}
\hskip-0.2cm
\epsfig{file=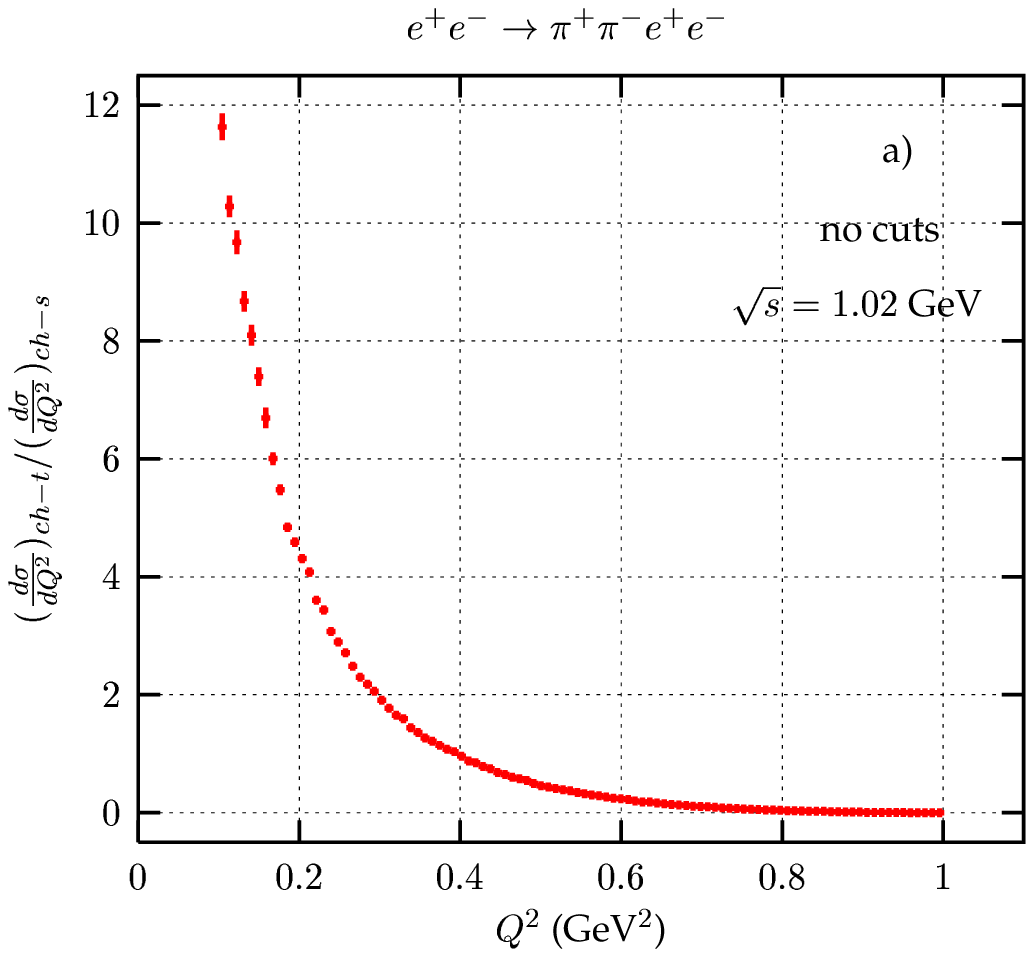,width=6.4cm,height=5.8cm}
\hskip-0.2cm
\epsfig{file=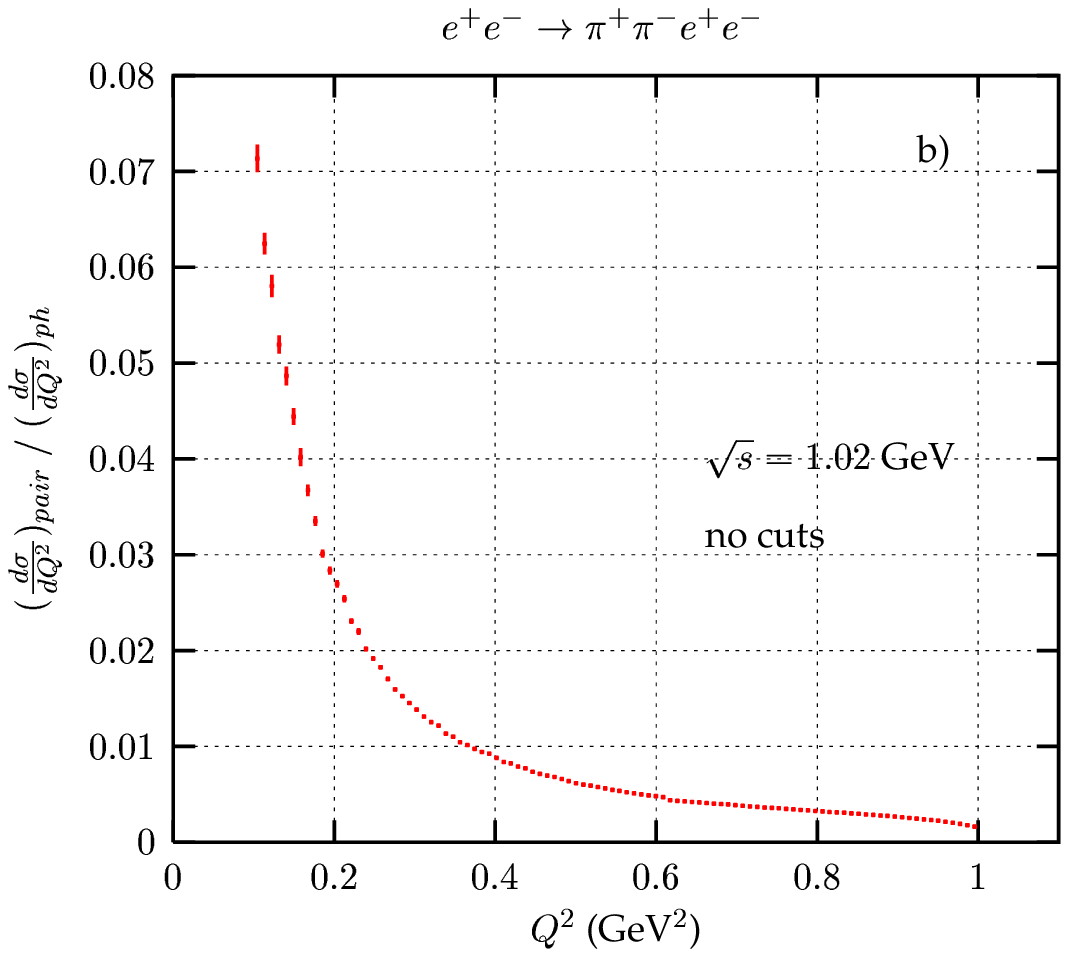,width=6.4cm,height=5.8cm}
\end{center}
\caption{(a) Ratio of differential cross sections of t-channel and s-channel 
of  the reaction $ e^+e^- \to \pi^+\pi^-e^+e^- $ (b)
Ratio of differential cross sections 
of  the  reactions $ e^+e^- \to \pi^+\pi^-e^+e^- $ (pair)
 and $ e^+e^-\to \pi^+\pi^- \gamma  $ (ph)} 
\label{przekroj} 
\end{figure}
\begin{figure}[ht]
\begin{center}
\hskip-0.2cm
\epsfig{file=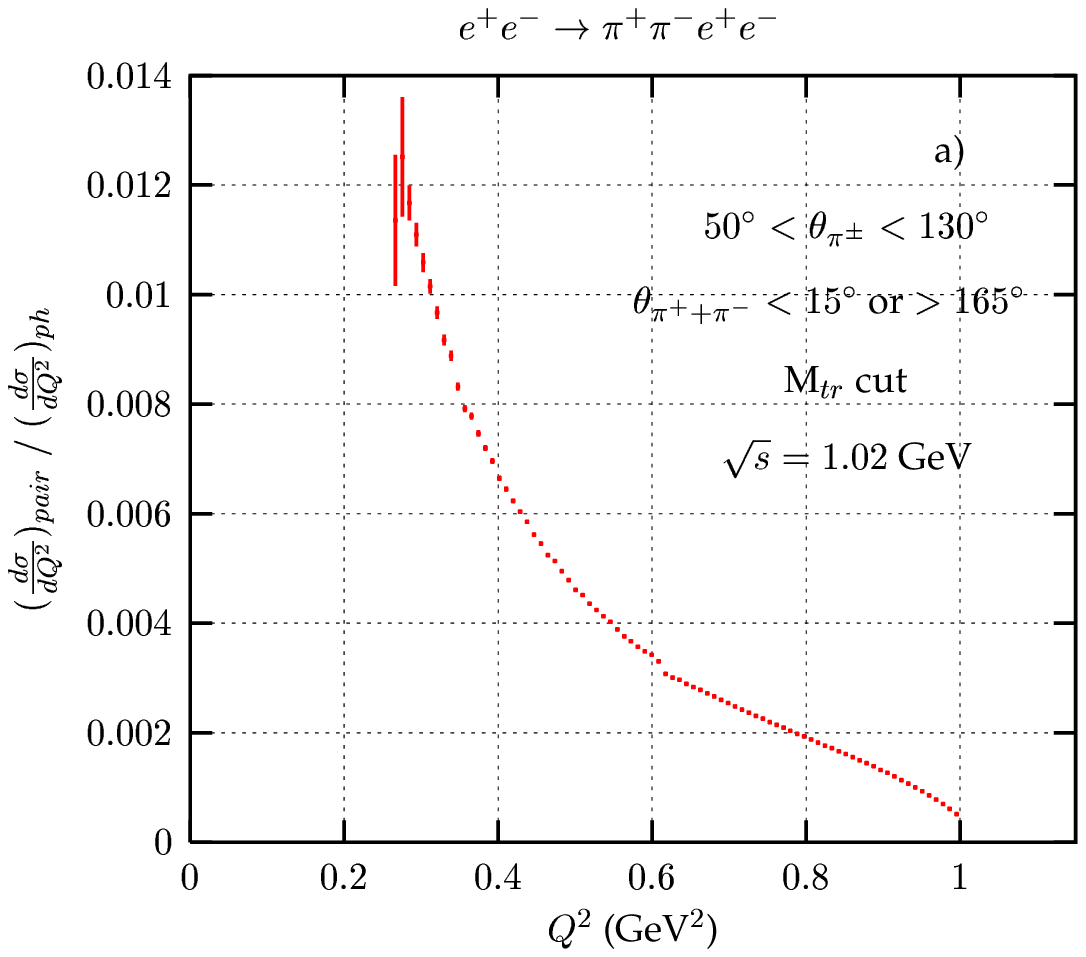,width=6.4cm,height=5.8cm}
\hskip-0.2cm
\epsfig{file=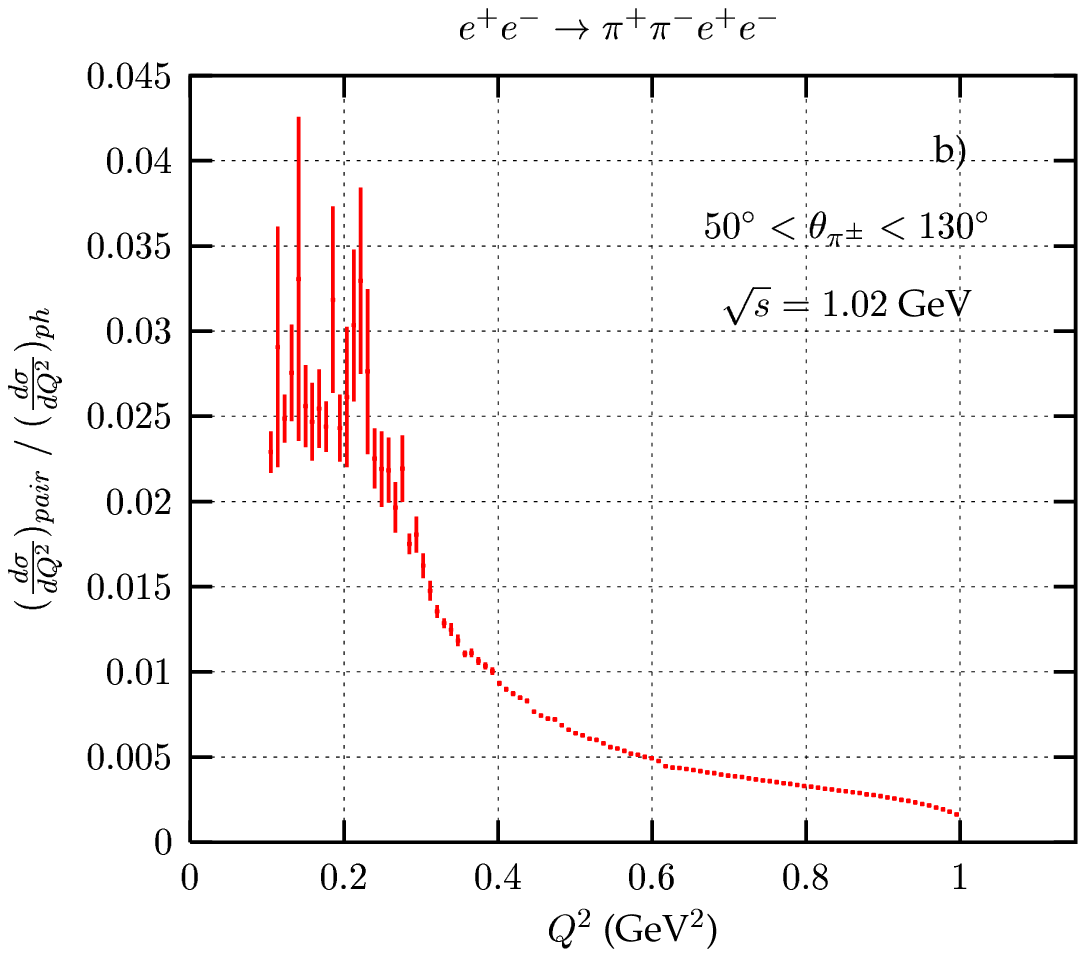,width=6.4cm,height=5.8cm}
\end{center}
\caption{
Ratio of differential cross sections 
of  the  process $ e^+e^- \to \pi^+\pi^-e^+e^- $ (pair)
 and $ e^+e^-\to \pi^+\pi^- \gamma  $ (ph) with two sets of cuts imposed
 (a) KLOE cuts (b) only pion angular cuts are imposed} 
\label{przekroj_cK} 
\end{figure}

All results from Monte Carlo simulation,
 presented in this section, 
  are for DA$\Phi$NE energy
1.02 GeV.
As observed already in the paper \cite{Hoefer:2001mx} t-channel diagrams 
contribute significantly in the small Q$^2$ region if no cuts are imposed.
That is shown in 
Fig.~\ref{przekroj}, where Fig.~\ref{przekroj}a presents the ratio 
of t-channel to s-channel contributions to
 the differential cross section of 
the reaction $ e^+e^- \to \pi^+\pi^-e^+e^- $ and Fig.~\ref{przekroj}b
 presents the ratio of the differential cross sections
 of the reactions $ e^+e^- \to \pi^+\pi^-e^+e^- $
 and $ e^+e^-\to \pi^+\pi^- \gamma  $.
 The t-channel diagrams contribution, if no cuts are imposed,
   is up to 12 times bigger than the s-channel contribution
 in the small Q$^2$ region
 and it is negligible for large Q$^2$ values.
\begin{figure}[ht]
\begin{center}
\hskip-0.2cm
\epsfig{file=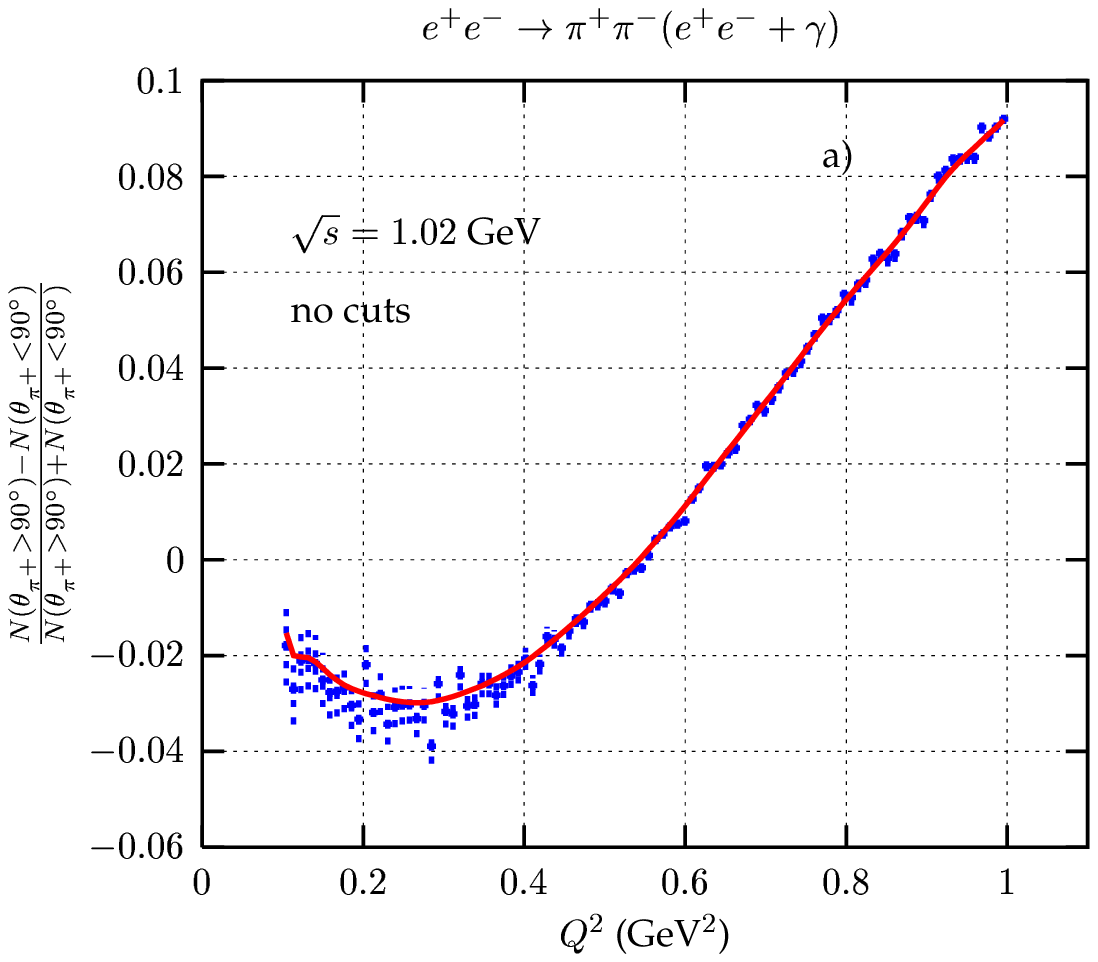,width=6.4cm,height=5.8cm}
\hskip-0.2cm
\epsfig{file=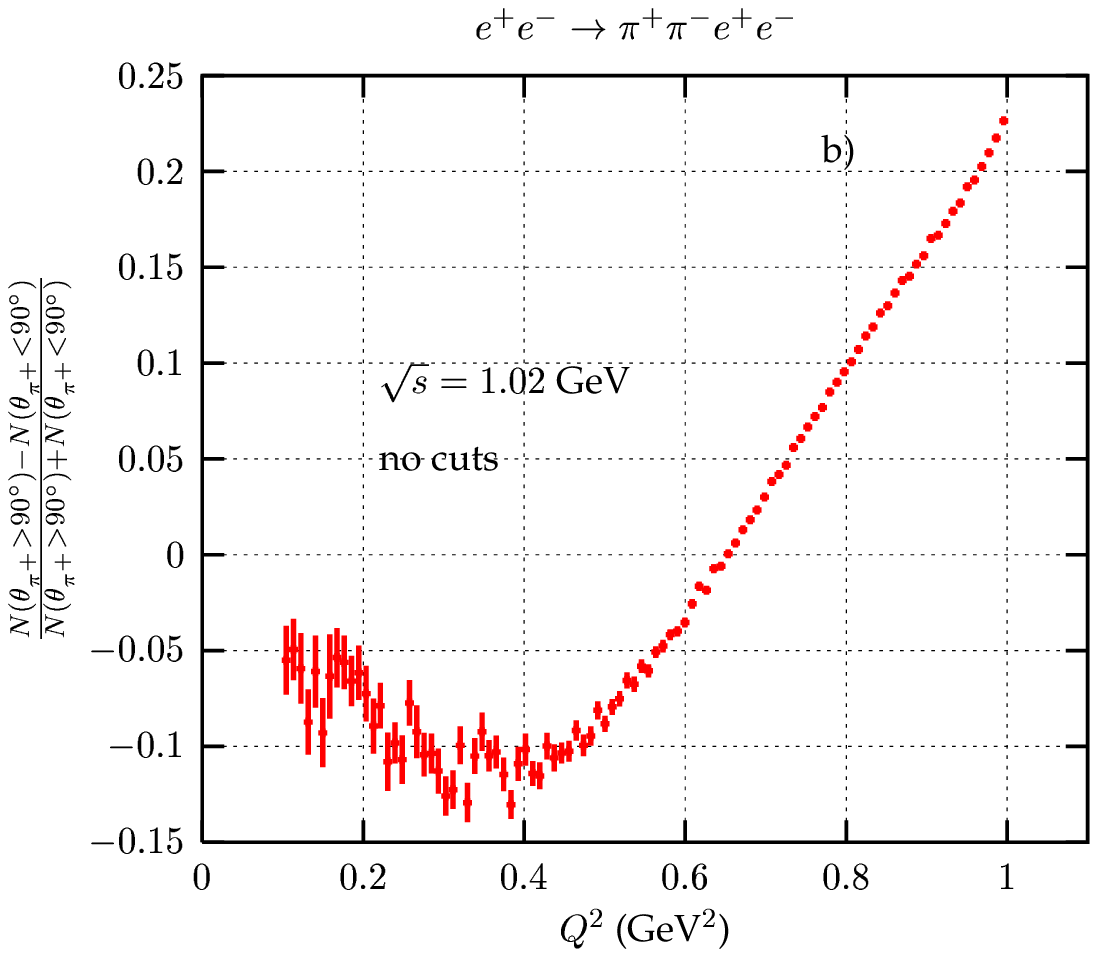,width=6.4cm,height=5.8cm}
\end{center}
\caption{Charge asymmetry $\frac{N(\theta_{\pi^+}>90^{\circ})
-N(\theta_{\pi^+}<90^{\circ})}{N(\theta_{\pi^+}>90^{\circ})
+N(\theta_{\pi^+}<90^{\circ})}$:
(a) for $ e^+e^- \to \pi^+\pi^-(\gamma + e^+e^-) $ (the solid line
 shows the charge asymmetry for $ e^+e^- \to \pi^+\pi^-\gamma$ only)
and (b) for $ e^+e^- \to \pi^+\pi^- e^+e^- $ (no cuts imposed)} 
\label{asymet} 
\end{figure}
\begin{figure}[ht]
\begin{center}
\hskip-0.2cm
\epsfig{file=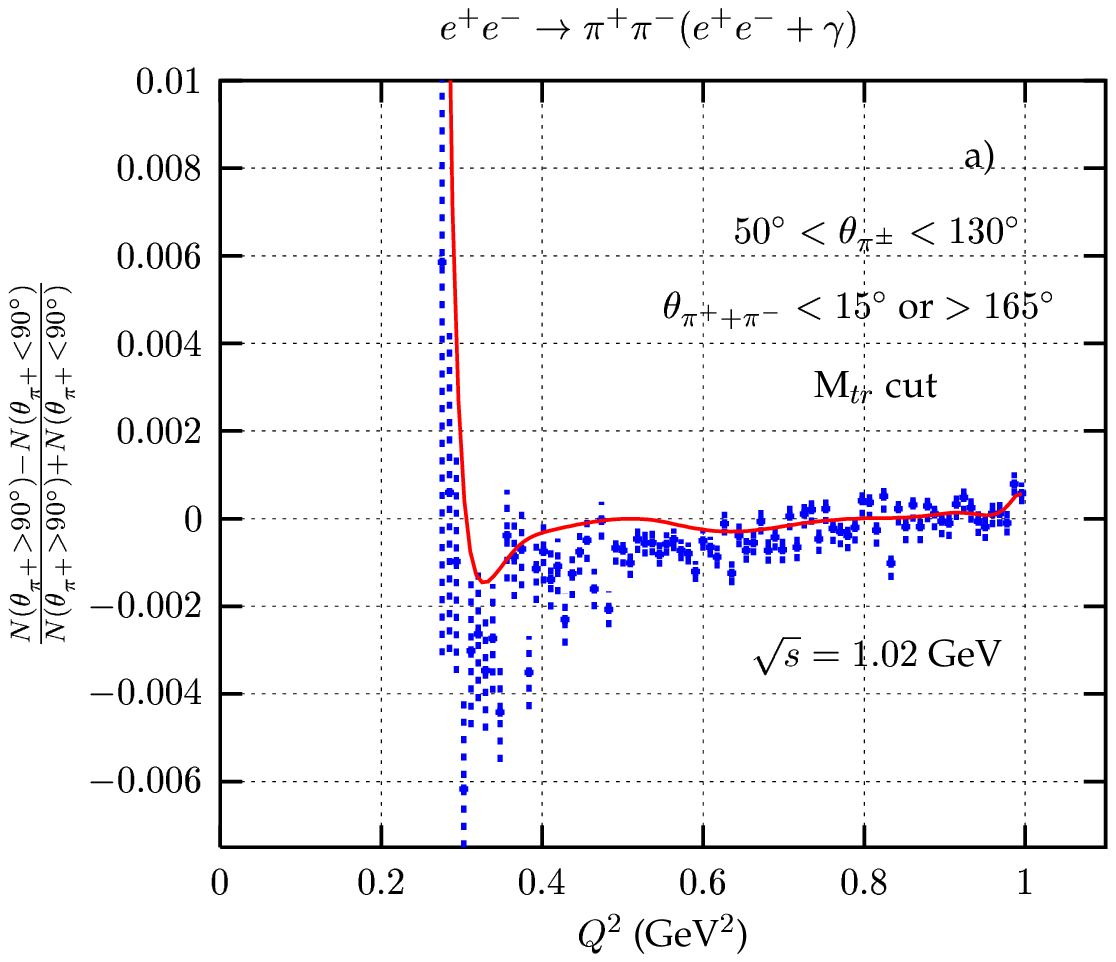,width=6.4cm,height=5.8cm}
\hskip-0.2cm
\epsfig{file=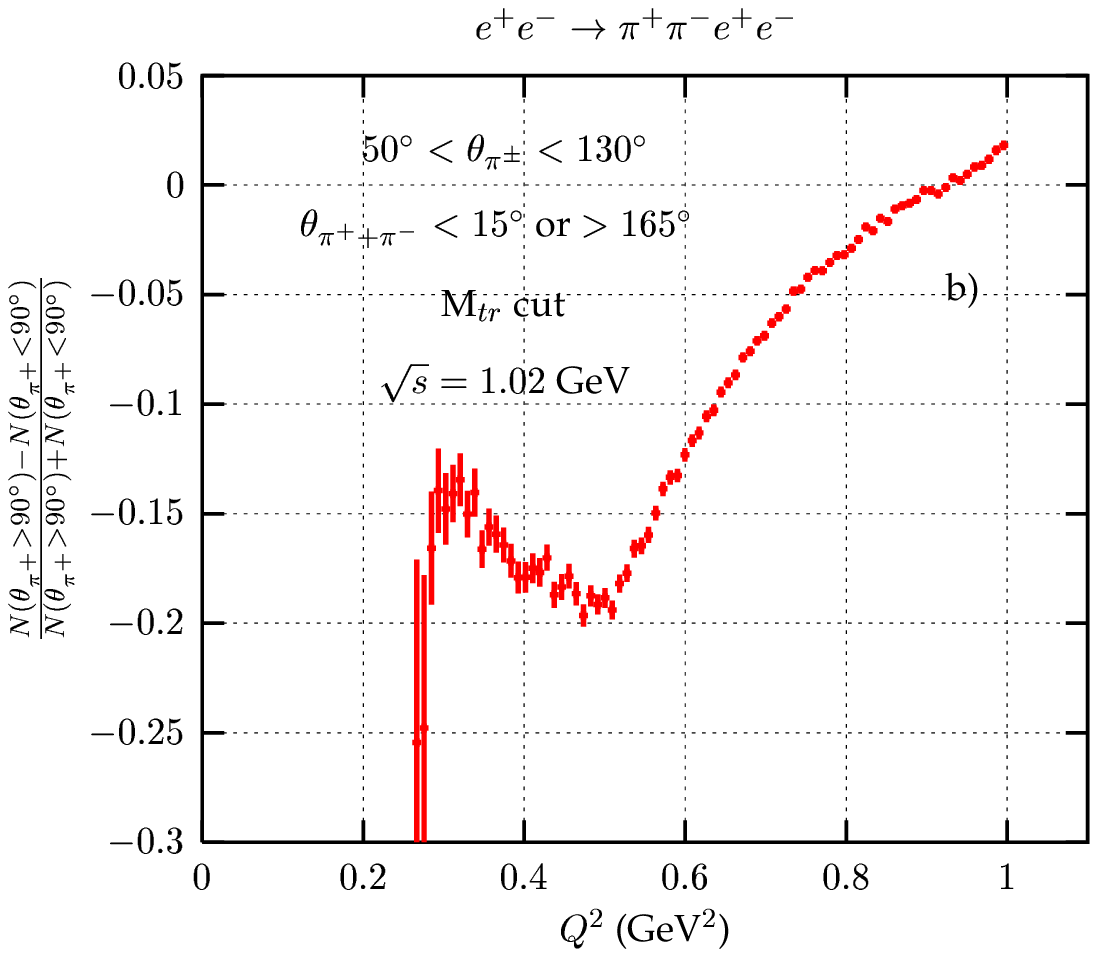,width=6.4cm,height=5.8cm}
\end{center}
\caption{Charge asymmetry $\frac{N(\theta_{\pi^+}>90^{\circ})
-N(\theta_{\pi^+}<90^{\circ})}{N(\theta_{\pi^+}>90^{\circ})
+N(\theta_{\pi^+}<90^{\circ})}$:
for (a) $ e^+e^- \to \pi^+\pi^-(\gamma + e^+e^-) $ (the solid line
 shows the charge asymmetry for $ e^+e^- \to \pi^+\pi^-\gamma$ only)
and (b) for $ e^+e^- \to \pi^+\pi^- e^+e^- $ (for KLOE event selection)} 
\label{asymet_cK} 
\end{figure}

The contribution from the pair emission
 to the photon emission depends however a lot on the event selection,
 even if the t--channel contribution still dominates at low Q$^2$.
It is seen from Fig.~\ref{przekroj_cK}, where for KLOE event selection
 (Fig.~\ref{przekroj_cK}a)
 one gets only up to 1.2\% contribution from the pair emission and 
 up to 2.5\% if one uses only the angular cuts on pions polar angles
(Fig.~\ref{przekroj_cK}b).
 The pair contribution at the $\rho$ peak is about 0.3\%
 for KLOE cuts and 0.5\% when only the angular cuts are imposed.
 Even if the pair contribution remains small it has to be taken into
 account if the aimed experimental precision is of the order of 1\%
 or better. 

Another important observable used to control the FSR contribution
 to the cross section of the reaction 
 $e^+e^- \to \pi^+\pi^- \ + \ {\rm photons}$ is 
the charge asymmetry. Again the measured value contains the pair 
 contributions if only pions are observed. However one can see in
 Figs.~\ref{asymet}a and  \ref{asymet_cK}a that this contribution is
 completely negligible both in the case where no event selection
 is used (Fig.~\ref{asymet}a) and in the case when KLOE event selection
 is applied (Fig.~\ref{asymet_cK}a). That conclusion holds
 despite the fact that the asymmetry for pair
 production itself is sizable in both cases (Figs.~\ref{asymet}b and
 \ref{asymet_cK}b).

\begin{figure}[h]
\begin{center}
\hskip-0.2cm
\epsfig{file=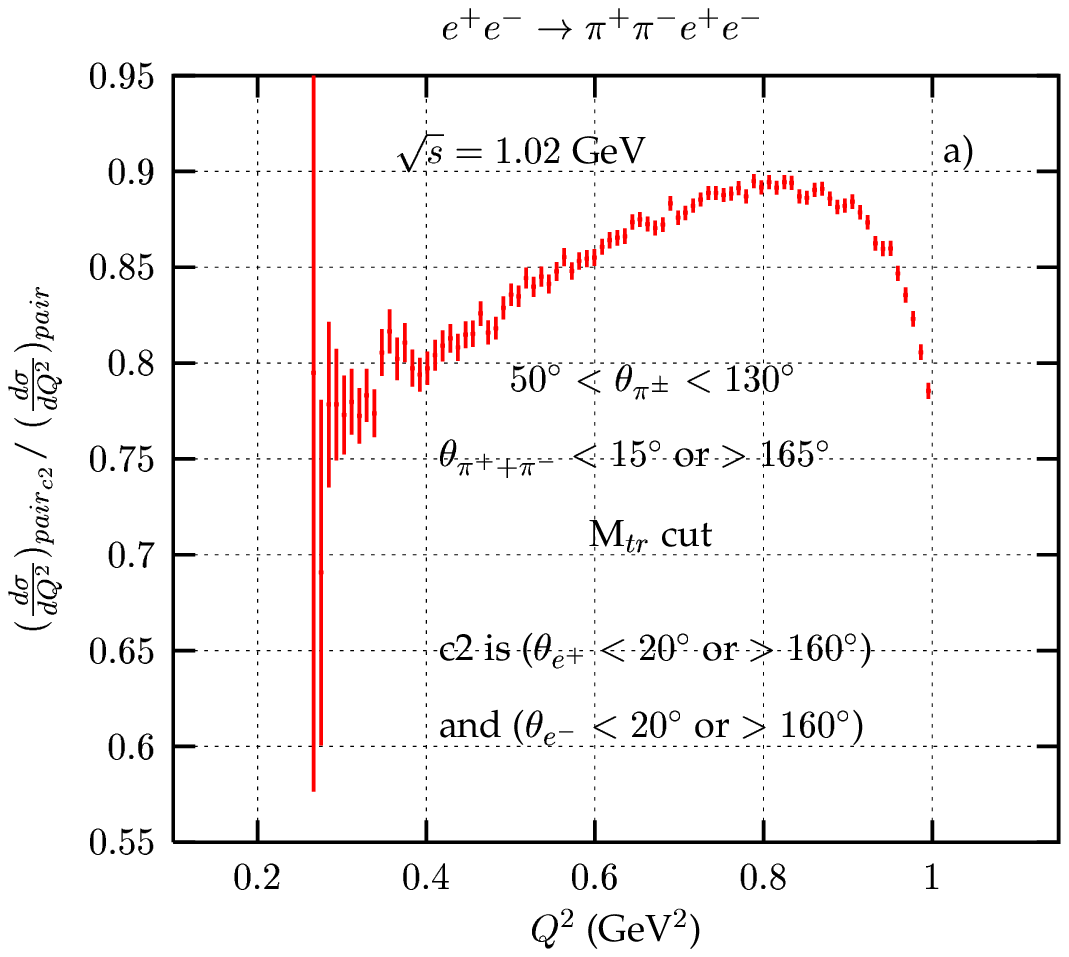,width=6.4cm,height=5.8cm}
\hskip-0.2cm
\epsfig{file=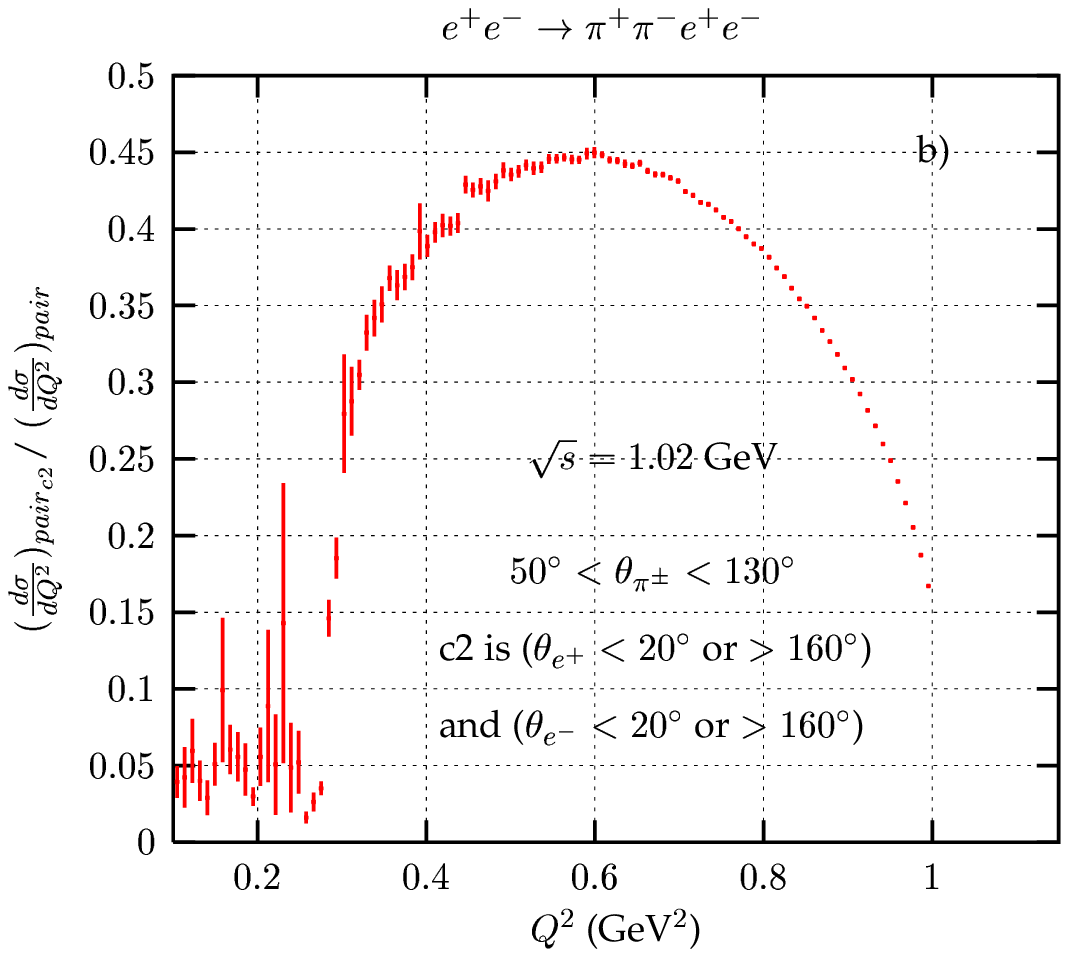,width=6.4cm,height=5.8cm}
\end{center}
\caption{Non-reducible pair production background } 
\label{veto_cK} 
\end{figure}

As stated already in \cite{actaHN},
 ISR and t-channel of electron pairs can be treated 
likewise ISR of photons, what results in the change of the radiator function 
in the radiative return method. Another possibility 
is to treat the pair production 
as a background to the process $e^+e^-\to\pi^+\pi^-\gamma(\gamma)$.
In that case one would like to reduce that background.
The way to do that is to veto the outgoing 
electrons and/or positrons.
However as shown 
 in Fig.~\ref{veto_cK}, where the percentage of the electrons and positrons
 escaping the detection ($ (\theta_{e^+} < 20^\circ $
 or $  \theta_{e^+} > 160^\circ) $
and $ (  \theta_{e^-} < 20^\circ $ or $  \theta_{e^-} > 160^\circ)  $)
is plotted, it is clear that it is practically impossible for KLOE
event selection (Fig.~\ref{veto_cK}a), as about 90\% of electron--positron
 pairs escapes detection. For looser cuts 
 one can reduce up to 50\% of the pair background (Fig.~\ref{veto_cK}b).

\section{Conclusions}

 A detailed analysis of the contribution of the reaction
 $ e^+e^- \to \pi^+\pi^- e^+e^- $
 to the $ e^+e^- \to \pi^+\pi^-$ cross section extraction
 via the radiative return method
 from the $ \sigma(e^+e^- \to \pi^+\pi^- + {\rm missing\  momentum})$
 measurement
 is presented basing on the Monte Carlo simulations with
 the EKHARA event generator.
 It is shown that t-channel diagrams contributions
dominate at small Q$^2$ values independently on the event selection.
The biggest possible contribution to  the 
 process $ e^+e^- \to \pi^+\pi^-\gamma$
reaches about 8 \% if no event selection is applied and it is reduced
   to 1.2 \% for KLOE event selection. 
The reaction
 $ e^+e^- \to \pi^+\pi^- e^+e^- $ does not contribute significantly 
to charge asymmetry measured in the reaction
 $ e^+e^- \to \pi^+\pi^- + {\rm missing\  momentum}$.

{\bf Acknowledgments}

The authors are grateful to W.~L.~van Neerven for providing the correct
 formula for the t-channel pair production \cite{bervanneerv}.


\end{document}